# Cartographic Relief Shading with Neural Networks

Bernhard Jenny, Magnus Heitzler, Dilpreet Singh, Marianna Farmakis-Serebryakova, Jeffery Chieh Liu and Lorenz Hurni

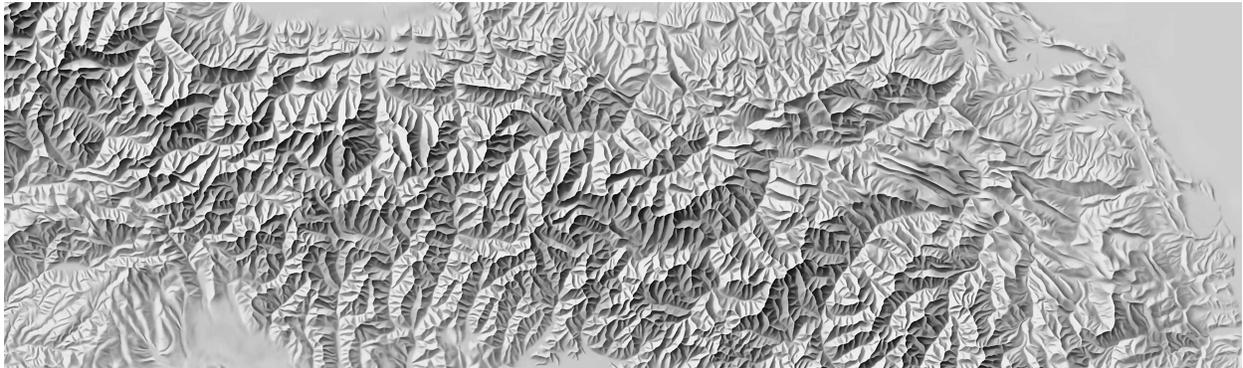

Fig. 1. Shaded relief of the Caucasus Mountains created with a neural network trained with a manual relief shading of Switzerland.

**Abstract**—Shaded relief is an effective method for visualising terrain on topographic maps, especially when the direction of illumination is adapted locally to emphasise individual terrain features. However, digital shading algorithms are unable to fully match the expressiveness of hand-crafted masterpieces, which are created through a laborious process by highly specialised cartographers. We replicate hand-drawn relief shading using U-Net neural networks. The deep neural networks are trained with manual shaded relief images of the Swiss topographic map series and terrain models of the same area. The networks generate shaded relief that closely resemble hand-drawn shaded relief art. The networks learn essential design principles from manual relief shading such as removing unnecessary terrain details, locally adjusting the illumination direction to accentuate individual terrain features, and varying brightness to emphasise larger landforms. Neural network shadings are generated from digital elevation models in a few seconds, and a study with 18 relief shading experts found that they are of high quality.

✦

## 1 Introduction

Cartographers commonly use shaded relief (or "hillshading") because it is an effective and aesthetic representation of terrain on maps. The main advantage of shaded relief is intuitiveness, as most map readers can easily and quickly interpret terrain elements ranging in size from large mountain ridges to small landforms, such as gullies or terraces. When drawing shaded relief manually, expert cartographers adjust the direction of illumination and modulate the brightness of individual terrain forms to visually accentuate the main terrain structures, while also making smaller, important details immediately recognisable [1]. A shaded relief is used in conjunction with other map elements, rather than as a standalone feature. It creates the impression of a three-dimensional surface on a flat, two-dimensional map and adds a pleasant aesthetic quality [2] (Fig. 2). However, creating a high-quality shaded relief requires specialised expertise and is immensely labour intensive, even when using digital image editing tools [3]. While standard algorithms shade digital elevation models in a fraction of a second, the resulting images lack expressiveness. The structure of terrain is more difficult to perceive with digital shading than with carefully crafted manual relief shading. For this reason, professional cartography often still uses manually produced shaded relief. For example, Switzerland's national mapping agency swisstopo, whose maps are frequently considered the gold standard in topographic mapping [4], scanned and georeferenced their manually shaded relief images when transitioning to digital production.

Our goal is to accelerate the production of high-quality shaded relief and make the process more accessible to map authors to better enable and encourage them to include shaded relief art in their maps. We were inspired by recent neural networks for the generation of realistic yet artificial graphical objects with image-to-image translation [5], neural style transfer [6], image completion [7], and colourisation of grey images (e.g. [8], [9]).

We create shadings from elevation models with a variation of a U-Net [10] (Fig. 1). U-Nets are a type of fully convolutional neural networks and were developed for image segmentation [11]. We employ a modified U-Net architecture for neural end-to-end rendering of digital elevation models. We train neural networks with high-quality manual shadings created by swisstopo cartographers. The neural networks are able to render shaded relief imagery that closely approximates the style and expressiveness of manual shading.

Our contributions include: (a) a neural network U-Net architecture for creating shaded relief imagery from digital elevation models; (b) insights on training neural networks with manual shadings; (c) methods for controlling illumination direction, level of detail, and height-dependent contrast of neural network shading; and (d) an evaluation of neural network shading with 18 experts, who provided very positive feedback and found that shaded relief created with our neural network method can reach the quality of manual shadings.

- Bernhard Jenny, Dilpreet Singh and Jeffery Chieh Liu are with Monash University, Melbourne. E-mail: {bernie.jenny, dilpreet.singh, jeffery.liu}@monash.edu.
- Magnus Heitzler, Marianna Farmakis-Serebryakova and Lorenz Hurni are with the Institute of Cartography and Geoinformation, ETH Zurich. E-mail: {hmagnus, mserebry, lhurni}@ethz.ch.

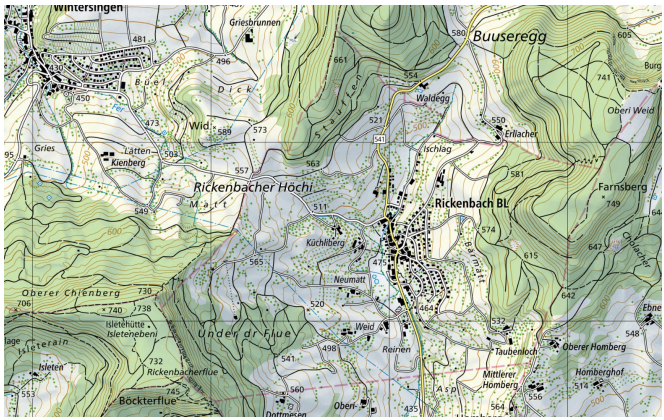

Fig. 2. Shaded relief with contour lines to portray the third dimension of terrain on a topographic map (1:50,000, reduced size, © swisstopo).

## 2 RELATED WORK

### 2.1 Design Principles for Cartographic Relief Shading

During the late 19th and 20th centuries, Swiss cartographers largely perfected the representation of terrain with shaded relief through continuous experimentation ([12], [13]). Eduard Imhof documented Swiss style relief shading in his book *Cartographic Relief Presentation* [1], which many contemporary cartographers consider a seminal contribution (e.g. [2], [12], [14]). The overarching goal of Swiss-style relief shading is to show the terrain as a three-dimensional surface by combining shaded relief with contour lines and optional colour subtly modulated with elevation and orientation of terrain [15] (Fig. 2). As described by Imhof [1], cartographers first delineate ridge and gully lines in mountainous areas, then start shading along the main drainage divides. Additional details are added gradually, beginning with dark shaded slopes then progressing towards the brightest illuminated peaks.

An oblique illumination from the upper left is preferred ([16]–[18]) and applied generally across the map to reduce the illusion of inverted terrain, where valleys may be seen as mountain ranges and vice versa ([19], [20]). However, the direction of illumination is adjusted locally to clearly accentuate the shape of individual landforms. Figure 3 illustrates where Imhof illuminated the mountain range in the upper left from the west instead of the standard north-western direction used elsewhere. Without this adjustment, both slopes of this major mountain would receive the same brightness and be indistinguishable from each other. The brightness of the main landforms is modified such that the shaded relief shows the main structures of the landscape. In Figure 3, area *A* is brightened, whereas area *B* is darkened to clearly show the two main sides of this mountain range, which is separated by a ridgeline between areas *A* and *B*. Imhof treated the neighbouring range to the east in the same way.

Depicting flat areas with a grey tone and excluding cast shadows are other common cartographic design principles for relief shading.

Aerial perspective, another important design principle, simulates the atmospheric scattering of light due to haze and other particles. The highest elevations, which are close to a map reader viewing the terrain from above, are shown with strong contrast between dark, shaded slopes and bright, illuminated slopes. Lowlands are more distant to the viewer and are shown with reduced contrast [21]. As a result, the contrast between illuminated and shaded slopes is accentuated along main ridges and is strongest along the highest mountain ridges and peaks, whereas grey values transition smoothly for rounded and lower landforms (Fig. 4).

In his book [1], Imhof dedicates considerable space to generalisation (that is, scale-adapted abstraction [22]) of shaded relief. At smaller map scales, less important or excessive details are removed to minimise their visual impact, while main ridges, valleys, and small but characteristic details are accentuated. Figure 5 shows how shaded relief is increasingly simplified at smaller scales.

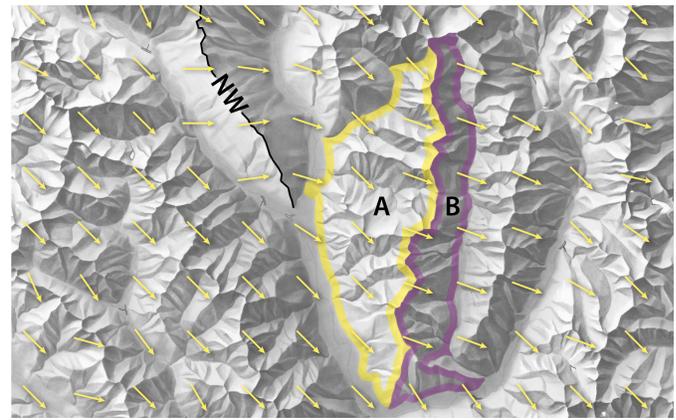

Fig. 3. Exemplary manual relief shading. Arrows indicate localised illumination direction. "NW" indicates a ridge oriented in the north-western direction with illumination adjusted to the west. Area *A* (yellow outline) is shaded in brighter tones and area *B* (purple outline) in darker tones to accentuate the ridge between *A* and *B*. (Ticino area by Eduard Imhof and Heinz Leuzinger, from shadedreliefarchive.com [23]).

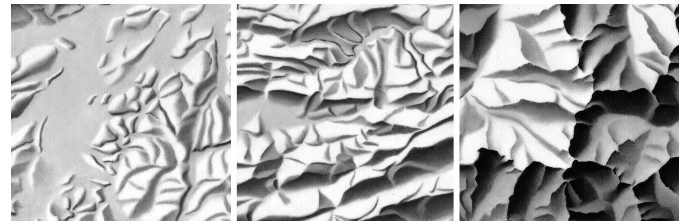

Fig. 4. Aerial perspective: contrast between illuminated and shaded slopes increases from lowland (left) to subalpine Jura Mountains (centre) and high alpine peaks (right, 1:500,000, © swisstopo).

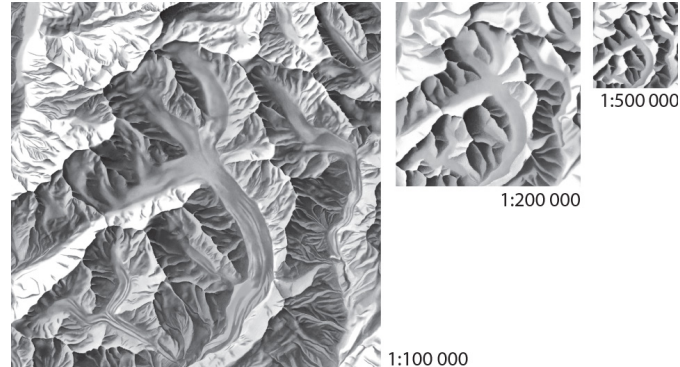

Fig. 5. Varying generalisation: sharp ridges are retained, but detail is reduced at small scales (Aletsch glacier, © swisstopo).

It is important to note that the design principles require the cartographer to read and interpret a landscape, usually from elevation contours and morphological break lines indicating main ridges and valleys. When a landscape is shaded by multiple cartographers, the resulting images will likely vary based on personal style and interpretation of the design principles (for an example of different styles, see [24]). Shaded relief is rarely produced manually nowadays because the process is an immense undertaking. Only a handful of specialists with the necessary training and artistic talent exist. For illustrations of their workflows, see ( [3], [25], [26]).

### 2.2 Digital Relief Shading and Illustrative Shading

**Digital terrain shading** of digital elevation models mainly uses variations of Lambert's diffuse reflection model. A large angular difference between the terrain normal and the illumination vector results in a dark value, while a small difference creates a bright value ([27]–[30]). To apply manual design principles to digital mapping, cartographers have developed algorithms for adjusting the illumination with the orientation of ridge and valley lines ([31]–[33]), combining multiple shadings with varying illumination directions ([34], [35]), applying sky illumination models ([36], [37]), and adjusting illumination with surface orientation [38], curvature [39], or slope and elevation [40]. A recent user study [41] compared different relief shading methods and found that no single shading method works well for all major types of landforms (e.g. alpine mountains, folded and eroded mountains, V and U-shaped valleys, etc.), but the study found that for many landforms, the 151 participants preferred the *clear sky shading* proposed by Kennelly and Stewart [36]. Besides cartographic shading algorithms, specialised generalisation operators exist for simplifying, smoothing, and enhancing features in elevation models that are applied before the shaded image is computed for maps at large scales (e.g. [42], [43]) and small scales (e.g. [24], [44], [45]).

**Illustrative visualisation** is a group of varied non-photorealistic rendering techniques that emphasise specific structural aspects of measured, simulated or modelled data (see Lawonn et al. [46] for a definition and review). The goal of illustrative visualisation is to communicate selected aspects more effectively than with conventional (photorealistic) shading and is often inspired by traditional illustration principles [46]. The cartographic relief shading algorithms described above represent a particular type of illustrative shading. *Exaggerated shading*, an illustrative shading technique proposed by Rusinkiewicz et al. [47], was even inspired by Imhof's principles for relief shading. They modify the diffuse shading equation for different frequency bands, emphasising surface details at various scales. *Light warping* by Vergne et al. [48] and *light collages* by Lee et al. [49] are two examples closely related to cartographic shading as they adjust the lighting direction to object geometry. Also related to our work is the simple *lit-sphere* shading-by-transfer method by Sloan et al. [50]. They use a shaded sphere as a lookup texture for shading.

**Neural rendering** is a varied and rapidly evolving field that combines generative machine learning with computer graphics. The overview by Tewari et al. [51] identifies diverse applications of neural rendering, for example, view synthesis, facial and body re-enactment, or photo relighting. A group of neural rendering techniques process computer graphics output (such as depth map, normal map, per-pixel material parameters, diffuse rendering, etc.) to improve low-quality renderings [52], simulate ambient occlusion ([53], [54]) or shading, and other illumination effects [54]. Our proposed method is inspired by these image-to-image translation operators; it transforms a scalar field with elevation values to a shaded image.

**Neural style transfer** is a popular type of neural rendering that uses a deep neural network to transfer the appearance of an image, building on concepts of content and style of images, which the inventors describe as follows [6]: Content is directly defined by the feature representations of each layer of a neural network. Style is defined for each layer by the feature correlation of the vectorised feature maps as expressed by the Gram matrix. Style and content are typically obtained from two different images that are fed into the same (pre-trained) neural network (e.g. VGG [55]). For transferring manual relief shading, the content image is a shading computed from an elevation model, while the style image is a manual shading of an arbitrary geographic area. To obtain the stylised shaded image, an image is initialised with random noise and then fed into the pre-trained neural network. Its pixels are iteratively optimised until losses in respect to style and content are minimised. Variations of this approach are discussed in two surveys ([56], [57]).

Tom Patterson, a renowned expert in terrain mapping, applied neural style transfer for relief shading [58]. First, he digitally shaded an elevation model, then used Prisma—a photo editing app—to apply an artistic style trained on paintings of rural life in India, which by happenstance, resulted in an improved, stylised shading. The neural style transfer was employed subtly for small changes. We experimented with neural style transfer and transferred hand-drawn relief shading to digital shadings computed from elevation models. While the manual style was closely replicated, we identified two problems: (1) neural style transfer could move or distort the geometry of terrain features, and (2) it occasionally omitted or invented terrain features that did not exist in reality. We were able to reduce these problems by tuning training parameters, but careful scrutiny of the resulting shading was required, and the problems could not be avoided entirely. We concur with Lawonn et al. [46] that neural style transfer does not lead to satisfactory results for illumination transfer.

**Summary:** Despite the many attempts to improve terrain shading since the early days of computer cartography, relief shading created with current methods does not reach the high quality of manual shadings. For example, existing methods for adjusting the direction of illumination are either cumbersome and complicated to use, require considerable manual intervention, or do not provide satisfactory results. Illustrative visualisation techniques have been applied to terrain models, and the development of some illustrative methods was even inspired by cartographic relief shading. However, there is no digital method capable of applying the design principles developed for manual shading in a coherent and practical way. The main problems are inconsistency in accentuating large landforms, lack of automated tools for locally adjusting illumination directions, and absence of high-quality generalisation of terrain models for relief shading, particularly at medium and small scales [58]. Neural style transfer is not applicable to cartographic terrain shading because mapping requires accurate positioning of all relevant terrain features and cannot accept the rendering of invented, non-existing features. Neural image-to-image rendering has proven successful in simulating illumination effects, but to our knowledge this has not been applied for transferring artistic relief shading for maps.

## 3 NEURAL NETWORKS FOR RELIEF SHADING

### 3.1 Network Architecture

Our neural network architecture is an image-to-image generator that is trained with a manually produced shaded relief image and a digital elevation model of the same geographic area. The trained neural network is then applied to a digital elevation model of any geographical area to create a shaded relief. The design style and amount of generalisation of the manual shading used for the training are transferred to the new shading. Consequently, the neural network replaces commonly used shading models, such as Lambert's diffuse reflection, and also removes excessive details. Neural network shading does not explicitly model illumination direction, but instead learns how to illuminate the model from the manual shading. The input consists of a scalar field with terrain elevation that is normalised between 0 and 1. No other terrain attributes that could be derived from the elevation model are provided to the neural network, such as normal vectors, slope steepness, or slope orientation.

When developing the network architecture, we experimented with a variety of convolutional neural network architectures. Unlike fully connected neural networks, convolutional neural networks can better exploit the spatial structure of images by training the weights of filter kernels rather than the weights used for linear combinations. We experimented with convolutional encoder-decoder networks as this family of architectures is well suited for transforming images, e.g. denoising and super-resolution [59] or colourisation ([8], [9]). We also experimented with state-of-the-art ResNet [60] and Inception [61] modules within our architectures. To guide the training of our models, we investigated various loss functions: L1, L2, structural similarity index (SSIM [62]), and perceptual loss [63].

Our final network is a U-Net [10]. U-Nets use connected encoding and decoding layers and have proven to be particularly useful for image segmentation, e.g., in the context of biomedical images [10] or geospatial raster data [64]. A U-Net consists of a contracting path (or encoder) to analyse the spatial context of an image and an expanding path (or decoder) that allows for accurate localisation of predicted features. Along the contracting path, the input image is gradually down-sampled using max pooling layers to obtain a compact multiscale feature representation. The expanding path makes use of up-sampling layers to gradually reobtain the dimensions of the input image. By copying the feature matrix of an intermediate step of the contracting path to the corresponding step of the expanding path, the network gradually combines low-level and high-level feature representations. Figure 6 shows an architectural schema of the U-Net variant used. The main differences between our architecture for neural network shading and the original U-Net architecture are as follows:

- The U-Net variant employs five instead of the standard four down-/up-sampling steps in each path. Our experiments have shown that other numbers of down-/up-sampling steps (e.g., four or six) also yielded satisfying results, but more than five steps did not result in visual improvements. Visual quality declined considerably with three or fewer steps.
- Dropout is used at each level of the network with increasing dropout rates towards the innermost layers. Figure 6 indicates dropout rates for each layer with dropout. The original U-Net architecture uses dropout only at the innermost layers. Figure 7 illustrates that dropout for each layer is essential; when dropout is only applied at the innermost level, the resulting shading shows heavy, local disturbance (Fig. 7, left). These artefacts are likely due to overfitting. Dropout effectively prevents overfitting by randomly supressing output from network units (Fig. 7, centre). This can be interpreted as training a multitude of different "thinned" architectures whose averaged result is approximated during inference [65], yielding the desired smooth shading.
- Weights are initialised via the method described by He et al. [66] instead of sampling from a normal distribution. Networks initialised from a normal distribution showed a higher probability of failing to learn certain desirable shading characteristics. For example, in Figure 7 (right), some slopes were depicted as plain grey areas without brightness modulation. This issue rarely occurs with He initialization, which is tailored to the initialization of ReLU layers and typically approaches better local optima than comparable methods [66].
- Power-of-two image tiles are used to avoid cropping operations required by the original U-Net architecture at intermediate steps. In Figure 6, the tile size is $256 \times 256$ pixels; the architecture can be adjusted to other power-of-two sizes.

As with the original U-Net architecture, ReLU layers [67] are used as activations after each convolutional layer. However, in contrast to segmentation where thresholding is applied to determine the class a pixel belongs to, no such operation is necessary for the proposed approach since the final predictions already represent the desired grayscale values. Clipping is applied for the rare case of a value exceeding the range [0, 1].

Because elevation models can be large, a terrain model is split into regular tiles and the network is trained to shade a single elevation tile. We typically use square terrain tiles with a width of 256 pixels or 512 elevation values. A shaded output tile is smaller than the corresponding input terrain tile. The goal is to allow the network to "look beyond" the area of the shaded output tile and adjust grayscale values with landforms that are larger than the size of the output tile. This enables the network to adjust the direction of illumination with the orientation of major mountain ridges that are not included in the output tile, and adjust brightness for major landforms. In the Figure 6 example, the cropped border is 50 pixels wide, resulting in output tiles of $156 \times 156$ pixels, covering 37% of the input terrain tile.

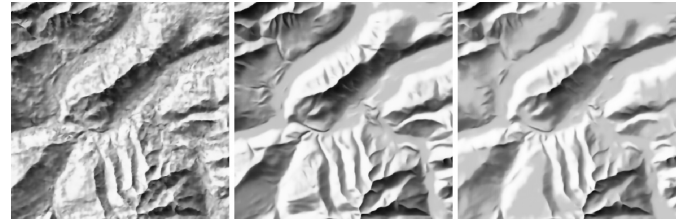

Fig. 7: The effect of dropout and He initialisation on shading quality. Left: Local noise occurs when dropout is only applied to the inner three layers. Centre: Noise is removed when dropout is applied to all five layers. Right: Initialisation from normal distribution (instead of He, as in the centre image) results in a lack of brightness modulation on slopes, which appear as flat plains.

When applying a trained network for shading an elevation model, the model is split into slightly overlapping tiles. The shaded output tiles are assembled into a final image using alpha blending to smooth differences in brightness along tile borders. An alternative to tiling is to modify the input and output layers of the network after training to render an entire elevation model in one go. It is possible to change the dimensions of the input and output layers because a U-Net only contains convolutional layers and therefore, only trains filter kernels, which do not depend on the dimensions of the input image. This alternative method is limited, however, by the size of available GPU memory.

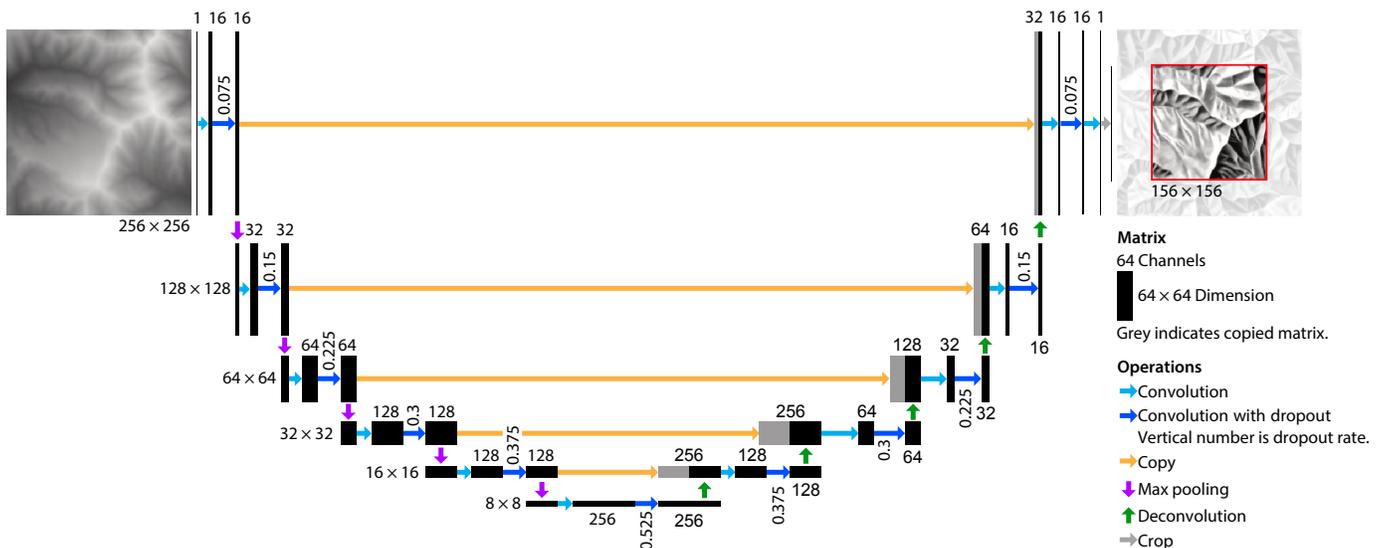

Fig. 6. The U-Net architecture for inferring a shaded relief tile (right) from a digital elevation model tile (left).

## 3.2 Data and Neural Network Training

The manual shaded relief used for training the neural networks was created for the Swiss national map series at scales between 1:25,000 and 1:1 million, of which the first map sheets were published in 1937. The multi-scale map series can be viewed online at https://map.geo.admin.ch. A small team of specialist cartographers at swisstopo created the shaded relief images mainly with airbrush tools ([68] [13]). A shaded relief image was created for each of the 247 map sheets at 1:25,000, 78 sheets at 1:50,000, 23 sheets at 1:100,000, and fewer sheets for smaller scales. The shadings are available as digital images that align with digital elevation models [69]. We trained neural networks with shadings at 1:50,000 and smaller scales. Cartographic generalisation during manual shading removed or strongly simplified small terrain features in maps at smaller scales. The shaded relief also shows local differences in style, contrast, and level of detail, which are the result of multiple cartographers contributing to this large work over many years. Consequently, the Swiss shaded reliefs contain many subjective decisions that were necessary to create this masterpiece.

We digitally edited the manual shadings before using them for training (Fig. 8). Most lakes had not been filled with a continuous grey tone in the manual shadings because lakes are typically shown without shading on the final maps. Using Adobe Photoshop, we individually selected the bright lake spots and filled them with a consistent grey tone. For example, in the lower part of Figure 8 (left), a north-facing mountain slope blends into the adjacent white lake. On Figure 8 (right), the lake has been filled with a grey tone creating a clear distinction between the lake and the bottom of the neighbouring slope. We also homogenised the grey tone for flat areas, which varied among map sheets and caused trained neural networks to render plains with spotty brightness. For example, the flat area in the lower right corner of Figure 8 (left) was corrected with the Photoshop brush tool because its brightness varied considerably. These manual corrections required roughly 25 hours of work for the 78 sheets at 1:50,000; however, it was only necessary to do one time before training the network.

Consistent and artefact-free shading of flat or nearly flat areas requires particular attention. In our training data, lakes and other flat areas exist at only a few altitudes (mainly between 200 and 570 meters in elevation). The networks initially produced random artefacts when asked to shade flat areas at some higher elevations. This is addressed by introducing artificial flat tiles placed at random elevations and corresponding shaded tiles with a continuous grey value. Alternatively, small sections of near-flat areas of the elevation model can be duplicated and then shifted *vertically* by a random amount. Training with shifted near-flat areas eliminates shading artefacts. When training the network, we also randomly shift the tiling origin of the input shading and input elevation model *horizontally* after a few epochs. Shifting the tiles reduces inconsistent spotty rendering of almost flat areas. Shifting every 25 epochs has proven successful.

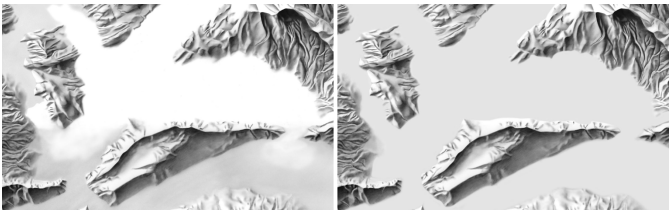

Fig. 8. Original manual shading (left) and shading for training with filled lakes and homogenised tone in flat areas (right, 1:50,000, Lake Lucerne, © swisstopo).

The loss function for training the network computed the mean squared error between the created shaded tile and the corresponding manual shading tile. It only considered the central area indicated by the red square in Figure 6.

The open-source TensorFlow library was used to model and train networks. Source images and terrain models for training always have the same number of pixels in width and height. They measured up to approximately 16,000 × 10,000 pixels. Training with an RTX 2080 GPU ran for 1,000 to 5,000 epochs, typically for one to five hours. We trained with a batch size of eight and used the Adam optimizer with values $\alpha = 0.001$, $\beta_1 = 0.9$, $\beta_2 = 0.999$, and $\epsilon = 10^{-8}$ as suggested by Kingma and Ba [70].

## 3.3 Controlling Illumination Direction, Contrast, and Generalisation

Three aspects for cartographic shading can be controlled with neural network shading: the direction of illumination, the contrast of shading according to terrain type, and the amount of generalisation. Our neural network architecture does not itself provide explicit control of these parameters. Instead, the input terrain model is manipulated—rotated, vertically shifted, or filtered—to give users control over these aspects.

The global lighting direction can be altered by rotating the terrain model before shading it and then rotating the resulting shaded relief image back in the opposite direction. This can be useful for creating a shaded relief with southern illumination, which is occasionally used to imitate the natural illumination direction in the northern hemisphere. Figure 9 shows an example shaded relief illuminated with standard north-western and alternative southern illumination.

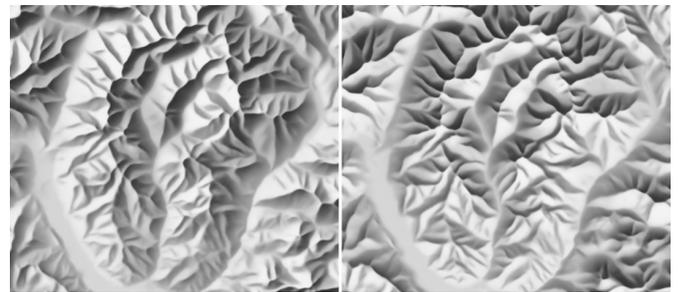

Fig. 9. Top-left (left) and bottom illumination (right, Sella Group, Italy).

In Switzerland, the topography is largely characterised by three varying landscape types. The first is plains with large lakes and rolling hills that have elevations between approximately 300 and 1,000 meters (m). The second type are the limestone Jura Mountains with deep river valleys and both round as well as pointy crests reaching approximately 1,500 m. Lastly, the Alps with characteristic pointed peaks and extremely steep slopes reach altitudes of more than 4,500 m. The manual shaded relief shows these three landscape types with increasing contrast between bright illuminated and dark shaded slopes (Fig. 4). We exploit the fact that the landscape types and contrast vary with elevation. Instead of using the full normalised elevation range between 0 and 1, the user can control the relative minimum and maximum. For example, when shading an alpine terrain, elevation values are normalised to the full range [0, 1]. However, when shading lowlands or rounded hills, the user can scale elevation values to [0, $k_{max}$] with $k_{max} < 1$. Figure 10 shows a shading of an almost flat lowland where $k_{max} = 20\%$: The lowland is shown with appropriate smooth brightness transitions throughout instead of high-contrast craggy edges. Controlling the lower limit can also be useful when shading high alpine terrain without lowland, that is, [$k_{min}$, 1] with $k_{min} > 0$. Controlling the elevation range allows for adjusting the shading style to fit the type of terrain, as well as shading adjacent terrain models with different landscape types (such as the lowlands and Himalayas in India) and then combining the shadings.

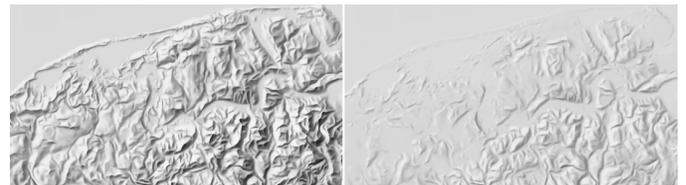

Fig. 10. Flat lowland shaded with full elevation range (0 to 100%) creates overly strong contrast and alpine appearance (left). When scaling elevations to 0 to 20%, the landscape rounded by glacial erosion is appropriately represented (right, northernmost part of Poland).

The neural networks can create generalised shading by removing distracting detail much like in manual shading (Figure 5). The amount of generalisation can be controlled by a pre-training and a post-training method. The pre-training method encodes the generalisation in the neural network. For example, we trained three networks with a strongly generalised manual shading at a scale of 1:500,000 (sections are shown in Fig. 4) and three terrain models with cell sizes of 200 m, 100 m, and 50 m. The network trained with a terrain model with 200 m cell size learned to retain and accentuate details because it saw only low-frequency terrain information during training. The network trained with 50 m cell size saw high-frequency information during training and learned to discard details accordingly. Figure 11 illustrates the increasing generalisation when the three networks shade the same elevation model with a cell size of 150 m.

The post-training method consists of down-sampling the resolution of the elevation model. Notably, sharp edges are retained, as networks learn to accentuate abrupt, high-frequency transitions (Fig. 12).

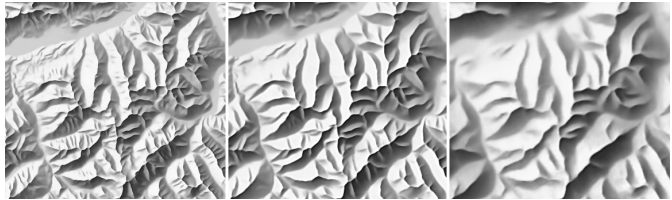

Fig. 11. Increasing generalisation by training networks with increasingly detailed terrain models (left: 200 m cell size, centre: 100 m, right: 50 m, Tyrol, Austria).

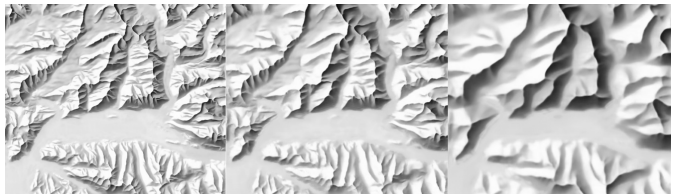

Fig. 12. Increasing generalisation by reducing cell size of terrain models post-training (30 m, 60 m, 120 m from left to right). All figures are rendered with the same neural network (Valdez, Alaska [71]).

## 4 EVALUATION

To evaluate the performance of neural network shading, we implemented an application for rendering arbitrarily sized terrain models. We used the hardware-accelerated Apple Core ML framework [72] and a 2019 laptop computer with an AMD Radeon Pro 5300M GPU. Rendering an elevation field with $5,000 \times 5,000$ values takes six seconds with an input terrain tile size of $256 \times 256$ pixels, an output tile size of $156 \times 156$ pixels, and 20 pixels alpha blending between neighbouring tiles. Tile-based rendering is scalable to large elevation models and is only limited by the amount of available memory.

As a testimony to the quality of shading with neural networks, swisstopo decided to use the presented U-Net architecture to extend the area covered by their existing manual relief shadings at 1:200,000 and 1:500,000 scales.

Evaluation of a shaded relief image is difficult to quantify with quantitative metrics because a single ground truth reference image does not exist. As a result, loss function values or similar metrics are not indicative of a network's performance. Visual qualitative inspection is the preferred method for evaluating shaded relief [32] and is also common for evaluating illustrative visualisation [46] and "creative" convolutional neural networks generating art or photorealistic renderings. Subjects for human evaluation of new rendering methods are the authors (e.g. [73], [74]), domain experts (e.g. [75], [76]), or general users on online survey platforms (e.g. [9], [77]). We first compare neural network shading to other digital shading methods and then report on an evaluation with relief shading experts who compared neural network shading to manual shadings.

### 4.1 Comparison with Digital Shading

Figure 13 shows a Lambertian diffuse shading, a neural network shading, and a clear sky shading of an area that includes plains, mountains with an intricate structure, northwest trending mountain ridges (which align with the main illumination direction), and a cone-shaped volcano. The neural network was trained with the 1:500k swisstopo shading (Fig. 4) with a cell size of 100 m, which resulted in a training image and a training elevation model of only $4,000 \times 2,700$ pixels (or approximately 150 tiles of $256 \times 256$ pixels).

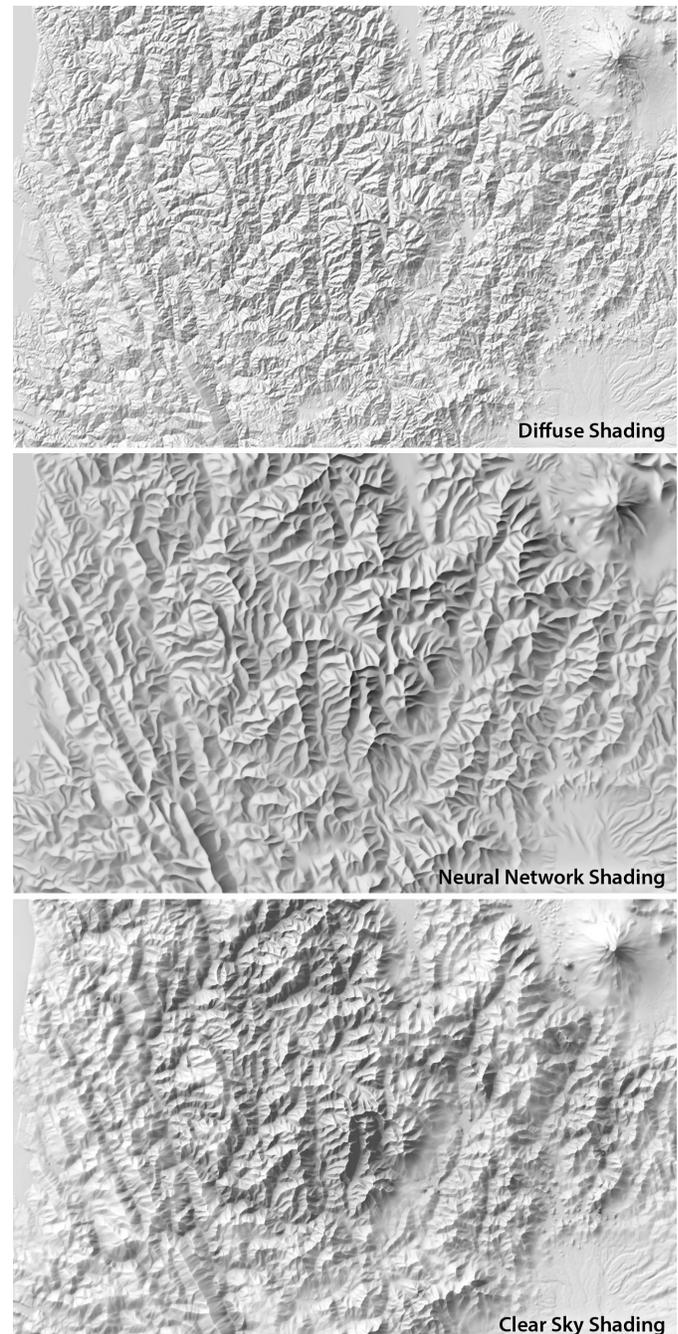

Fig. 13. Standard diffuse shading (top), neural network shading (centre), and clear sky shading (bottom). Northern California, elevation model with 180 m cell size, filtered for clear sky shading.

The three shadings in Figure 13 were created with an elevation model with a cell size of 180 m. The neural network shading shows terrain features at correct locations, which is relevant for aligning the shading with other map information, such as drainage networks and summit points. The network did not invent non-existing features or omit important features. The neural network successfully learned to replicate the design principles for terrain shading: the network clearly depicted the main landforms, accentuated high elevations with stronger contrast than low valleys, and locally adjusted the light direction. Illumination is rotated to the west for the northwest trending hills in the lower left corner. The same hills are difficult to perceive in the diffuse shading because the illumination direction aligns with the hill crests. On the diffuse shading, the rendition of the amalgamation of mountains at the centre is confusing and cluttered; the neural network shading more clearly shows the intricate shape of this terrain. The neural network also successfully generalised the terrain; it removed excessive details and accentuated main ridges and valleys, resulting in a terrain rendering that is easier to read and combine with other map information than the diffuse shading.

Clear sky shading [36] is a variation of ambient occlusion shading that was preferred over other relief shading methods in a recent user study [41]. Before rendering the clear sky shading with the SkyLum software [78], we removed details from the elevation model in mountainous areas with line-integral convolution filtering [43], which retains edgy ridges. Clear sky shadings are considerably darker than diffuse and neural network shadings when applied with a recommended fivefold vertical exaggeration [78]. In Figure 13, brightness was increased to approximate the appearance of the other shadings. Clear sky shading creates dark values for narrow valleys and bright lines for sharp ridges. It accentuates and exaggerates small details, resulting in an uneven noisy distribution of bright and dark areas; grey values mainly vary with narrowness of valleys. In comparison, neural network shading varies values according to the relevance and elevation of mountain ridges. Diffuse shading and clear sky shading do not adjust illumination for ridges and other landforms that align with the direction of illumination, as in the lower left corner of Figure 13.

## 4.2 Comparison with Manual Shading by Experts

An online survey collected feedback from expert cartographers to determine how they rate shadings created with neural networks, whether they prefer manual shading or neural network shading, and identify shortcomings of neural network shading. This study did not evaluate whether mountain ridges or other terrain features were easier to perceive in manual or neural network shading than with other digital shading methods.

**Participants**: Eighteen experts (4 female, 14 male) participated; names are listed in the Acknowledgments section. Sixteen have extensive experience in creating high-quality shaded relief either manually or with digital tools, and two have published research results aiming at the digital automation of design principles for relief shading.

**Setup**: We asked experts to rate and comment on ten pairs of shadings (Fig. 14) using an online survey. The two images in each pair showed shadings of the same geographic extent; one image was generated with a neural network and the other was drawn manually. The order of the two images within each pair was random and participants were not informed how the shadings were created. The two images were displayed side-by-side (except for the first pair, which was arranged vertically) with a total width of 1,000 pixels.

**Tasks**: For each pair, participants were asked to comment on strengths and weaknesses of both images and rate each image on a 5-point Likert scale between poor and excellent. For each pair, participants also indicated which image was better for a topographic map, which image was more aesthetic, and which image was created manually. These questions were answered with one 5-point Likert scale per image pair. The possible answers were *clearly left (1), rather left (2), no preference/I don't know (3), rather right (4), and clearly right (5)*. For reporting the results, we converted the answers for preference and aesthetics from numerical Likert values to corresponding categories: 1 = *clearly manual*, 2 = *manual*, 3 = *no preference*, 4 = *neural*, 5 = *clearly neural*. Answers to the question asking users to identify the manual shading were converted to 1 = *clearly wrong*, 2 = *wrong*, 3 = *uncertain*, 4 = *correct*, 5 = *clearly correct*.

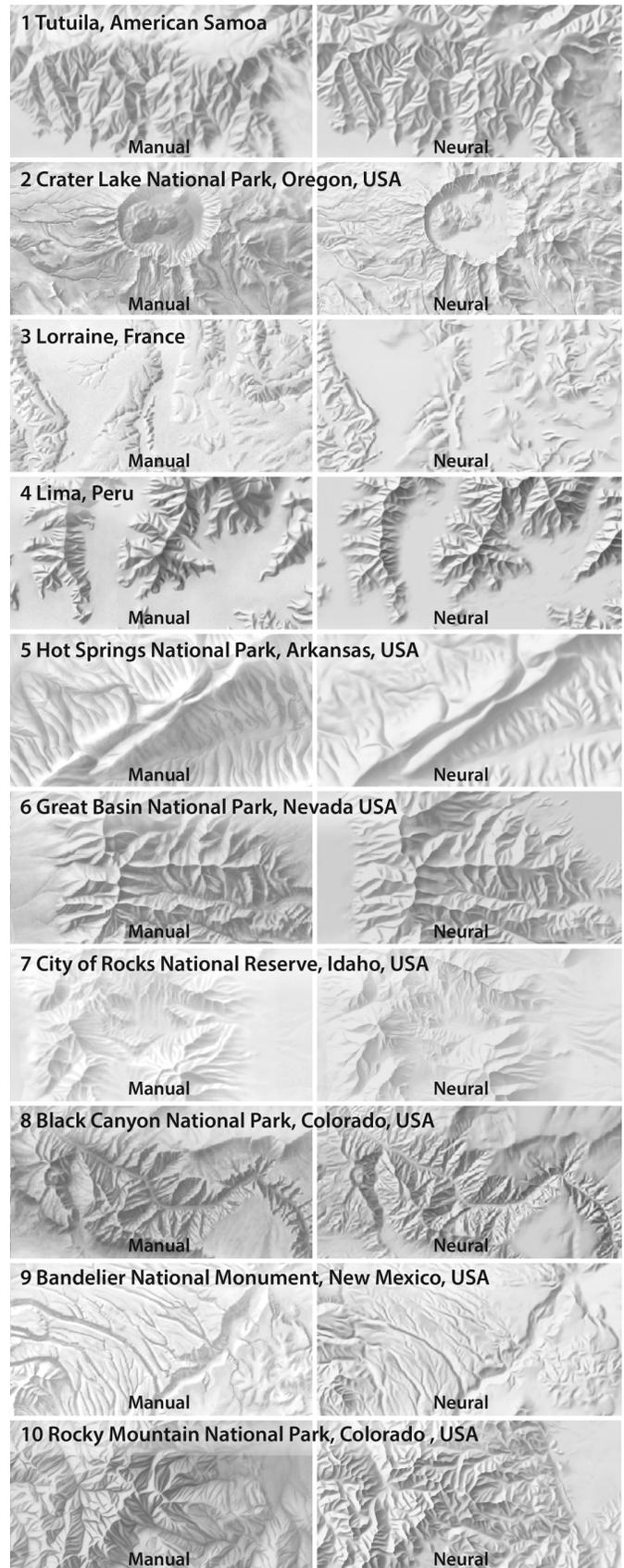

Fig. 14. Sections of the ten pairs of manual and neural network shadings for the expert evaluation. Full resolution images are included in the supplementary materials (manual shadings of #3 and #4 © Swiss Conference of Cantonal Ministers of Education (EDK), Swiss World Atlas).

**Data**: We selected high-quality manual shadings of geographic areas outside of Switzerland because the neural networks could have been overfitted and learned to replicate Swiss geography instead of creating generic relief shading. Our goal was to assess the applicability of neural network shading to other areas. We initially compiled 17 pairs of shadings and had two experts take the survey. We then removed the seven pairs that received the lowest scores for the manual shadings from the two experts in order to avoid comparing our neural network shadings to suboptimal manual shadings. The final set included shaded reliefs at various scales and terrain types. Seven manual shadings were created by the U.S. National Park Service for their park brochures (Fig. 14, #2 and #5–10), two shadings (#3 *Lorraine* and #4 *Lima*) were from the Swiss World Atlas [79], and #1 *Tutuila, American Samoa* was created for another atlas [80]. Most shadings are available through the Shaded Relief Archive [23]. The pairs of shadings are included in the supplementary materials. For each manual shading, a neural network shading with a similar level of detail and overall appearance was created by controlling contrast and generalisation with the techniques described in Section 3.3. The default upper-left illumination direction was used for all shadings. Global brightness and contrast of neural network shadings were adjusted such that both images of a pair had a similar appearance.

**Results**: Experts spent considerable time answering the ten questions; the average time was very close to one hour (excluding five outliers who reported longer intermissions). Figure 15 presents the expert ratings. The first column shows overall ratings for each shading ordered by pairs of manual and neural network shadings. While the experts agreed that most manual shadings that we selected were of *fair*, *good*, *very good*, or *excellent* quality, we were nevertheless surprised by how critically they rated the manual shadings. The manual shading of pair #10 *Rocky Mountain National Park* was an outlier, because it received five *poor* and four *fair* ratings. Some experts criticised its emphasis on the drainage network, lack of accuracy, and the stylised representation.

Ratings summed across all neural network shadings were: 15% *poor* or *fair*, 32% *good*, 43% *very good*, and 10% *excellent*. Most neural network shadings received similar ratings as their paired manual shadings, particularly maps showing mountain ranges resembling Swiss topography (#4, 6, 7), and/or maps at large scales (#1, 4). The neural network shading of pair #3 *Lorraine*, #5 *Hot Springs*, and #8 *Black Canyon* received more than two *poor* or *fair* ratings.

When asked to indicate "which of the two shaded reliefs is better for a topographic map", there were five pairs with the neural network shading preferred (#1, 2, 4, 7, 10), one pair with the manual shading preferred (#3), and four pairs without a clear preference (#5, 6, 8, 9). Experts preferred the neural network shading when they also rated the neural network shading higher (#1, 2, 4, 7, 10) and vice versa (#3). A recurring theme in comments was the more consistent generalisation of neural network shadings, which the experts appreciated. An example is the #5 *Hot Springs* pair: the manual shading received many *fair* ratings due to the less than perfect selection and representation of the multitude of small details.

Aesthetic preference ratings aligned with general preference ratings (compare the pairs of bars in the right column of Fig. 15). Across many pairs, experts commented positively on the more balanced, clear, and consistent appearance of neural network shading. For example, for the manual shading of #2 *Crater Lake*, some experts criticised the dark, "globby" and inconsistent look. At the same time some experts also liked when manual shadings highlighted landforms that were of particular interest, for example, the crater in #2 *Crater Lake* or the valley bottom of #8 *Black Canyon*.

#3 *Lorraine* is the only pair of shadings where experts preferred the manual shading to the neural network shading. They commented that they liked the clear hierarchy of the large major landforms in the manual shading and criticised the lack of hierarchy and indifferent representation of the landforms in the neural network shading.

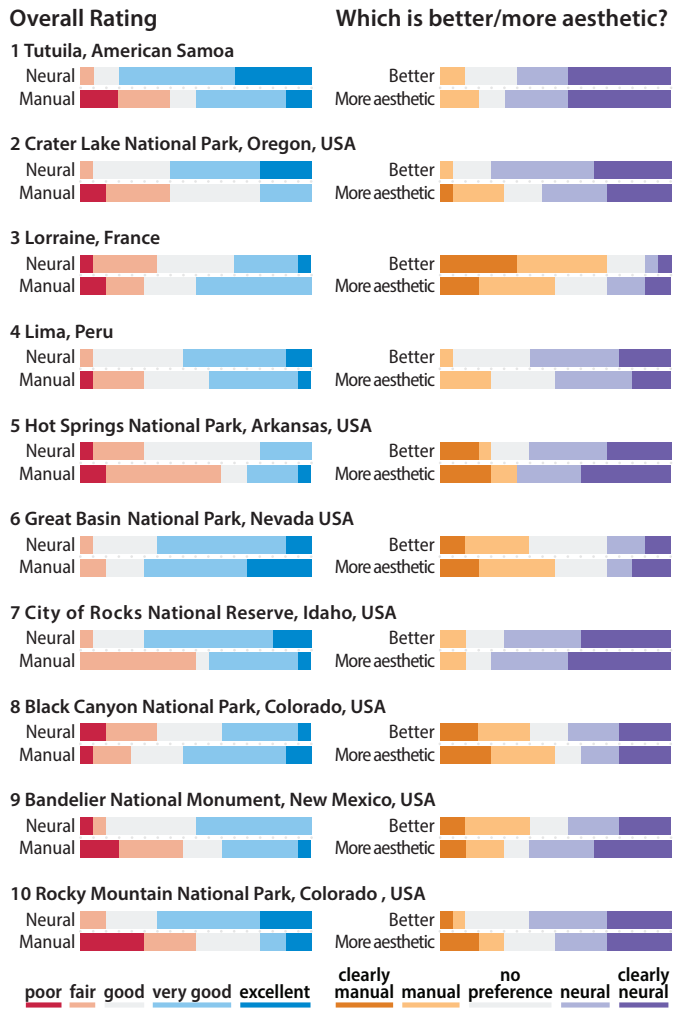

Fig. 15. Likert scale ratings (left), preference for a topographic map and aesthetics rating (right) by 18 experts on ten pairs of manual and neural network shadings.

Experts had difficulty identifying the manual shading for some pairs. For each pair, they answered the question "which of the two shaded relief images was created manually?" on a 1–5 scale where 1 = "clearly left", 2 = "rather left", 3 = "I don't know", 4 = "rather right", and 5 = "clearly right" ("left" and "right" referred to the two randomly ordered images). Figure 16 shows that the manual shadings that were most difficult to identify were #1 *Tutuila*, #3 *Lorraine*, and #7 *City of Rocks*. The manual shading in #10 *Rocky Mountain* was most easily identified.

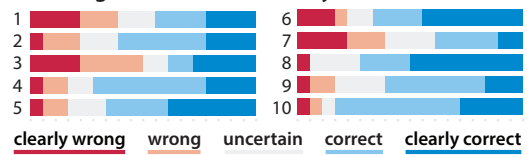

Fig. 16. Identification of manual shadings by 18 experts (numbers as in Figure 15).

The survey also asked experts how likely they would use neural networks for relief shading. Fourteen replied they were very likely or very interested in using it, one replied with "possibly", one was unlikely to use it, one is retired and no longer creates maps, and one did not provide an answer. Experts could also provide optional comments, some of which were enthusiastic in support of neural shading.

## 5 DISCUSSION

The proposed U-Net architecture for image-to-image neural rendering successfully shades elevation models with the expressive Swiss-style relief shading. In most tested examples, the neural network shaded relief images accurately portray major landforms as well as small, relevant terrain features. We did not detect any invented, non-existing features in our test shadings.

Compared to the clear sky shading, our neural network test shadings show terrain with a more even grey value distribution and consistent generalisation. For example, the neural network shading in Figure 13 locally adjusts the light direction to show all major landforms, resulting in a clearly structured rendering.

The 18 relief shading experts generally rated neural network shadings and the corresponding manual shadings similarly in overall quality, preference, and aesthetics. Only one of ten neural network shadings was rated lower than the corresponding manual shading. For many pairs, experts had difficulty identifying the manual shading from the neural network shading. Experts also liked the aesthetics of neural network shading and most commented positively about it.

The current neural network shading has limitations. There is an occasional lack of detail in flatter areas, when a neural network is applied to a terrain model that has a markedly different cell size than the model used for training. In Figure 17 for example, a neural network trained with an elevation model with a cell size of 200 m is unable to accurately render a relatively flat elevation model with a cell size of 5 m (left). Details in the terrain are difficult to discern due to the lack of light modulation in flat areas. An elevation model with a cell size of 10 m results in fewer artefacts (centre), and with a cell size of 20 m (right), the terrain is accurately rendered albeit with stronger generalisation. This particular example neural network cannot handle elevation models with a cell size that is more than 10 times smaller than the cell size of the elevation model used for training.

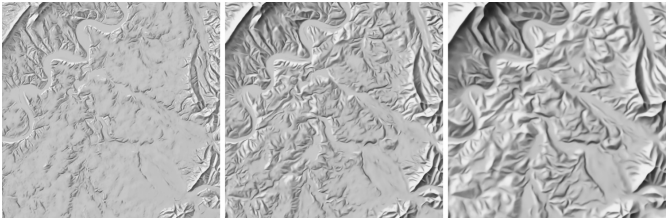

Fig. 17. Poorly rendered flat areas (left and centre) when the rendering cell size (5 m, 10 m and 20 m from left to right, respectively) is drastically smaller than the training cell size (200 m, Massanutten Mountain, USA [71]).

Some experts also criticised a shortage of details in flat areas in neural network shadings. For example, for #3 *Lorraine* (Fig. 14), many experts commented that the characteristic flat-topped terrain features on the left side did not show up well and that the terrain features were not depicted with sufficient differentiation and seemed to be unconnected. We expect the reason for these shortcomings is that the manual shaded relief used for training the network did not contain similar features and was at a larger scale. In our tests, trained networks were able to successfully shade landforms that do not exist in Switzerland, such as a volcano (Fig. 14, #2 *Crater Lake*) or high plateaus and canyons (#8 *Black Canyon*). However, further testing is required to ascertain whether every conceivable landform type can be depicted adequately.

Additionally, the networks are unable to locally adjust brightness to highlight specific features that are particularly relevant for a map. For example, the crater and canyon bottoms in #2 and #8 (Fig. 14) are purposefully depicted with dark values on the manual shadings to emphasize these landforms, while the neural network shadings use a generic grey tone for all flat areas.

We initially thought very large manual shading images covering many map sheets would be needed for training the neural networks, but found that relatively small shadings were sufficient. For example, the network for creating the shading in Figure 13 (centre) was trained with a shaded relief image of a single map sheet with 4,000 × 2,700 pixels. A constant tone needs to be applied evenly to all flat plains, otherwise speckles appear in the shading of flat areas; lakes and other waterbodies need to be shaded like plains or the trained network invents phantom water features; and contour lines, road networks, or other map features cannot be included in the training shading because the network will render these features at random locations.

Limiting the replicability of this research is the random initialisation of the network weights, which can influence the visual appearance of renderings. It is possible that two trainings with identical parameters and input data, but random initialisation produce networks that will create slightly different shadings. We found that He initialisation reduces this issue; however, we occasionally encountered networks that produced disturbing artefacts that disappeared when the network was trained again with a different set of random initial network weights.

## 6 FUTURE WORK AND CONCLUSION

Our example renderings indicate that the neural networks learned to apply established design principles for manual cartographic relief shading including locally adjusting brightness and contrast to accentuate important terrain features. In our tests, the neural networks also successfully generalised the shadings by removing excessive terrain detail while retaining and accentuating important landform elements such as mountain ridges or valleys.

Users can adjust the contrast and appearance of the neural network shading according to the terrain type (between flat lowlands and alpine peaks) by adjusting the relative elevation range (see Section 3.2). We aim to automatically adjust the relative elevation range based on the terrain type. This would both eliminate required user input and enable the creation of detailed global neural network shadings for web maps.

We would also like to improve the rendering of flat areas and sharp ridges, which occasionally appear slightly blurred when compared to manual shading. The networks are not always successful if the cell size or terrain type differs considerably from the training data (Fig. 17). We plan to develop neural networks that are more robust across a wider range of possible cell sizes. Recently introduced neural networks architectures, with more sophisticated connections such as stacked hourglasses [81] or HRNet [82] may help achieve these goals. We plan to extend our work to coloured relief shading by training networks on manual shadings that vary colour with elevation, exposition to illumination, and artistic interpretation of landforms.

With our application of deep learning to artistic relief shading, we hope to contribute to the curation of this cartographic expert knowledge and bring high-quality shading to the reach of modern mapmakers. We believe it is promising to explore the application of neural network rendering to other types of information visualisations and artistic illustration. While we optimised training and model parameters for a particular set of landforms and a specific style of cartographic relief shading, it remains to be seen how the proposed network architecture and training setup may be applied to other visualisations.

## ACKNOWLEDGMENTS

We thank the National Geographic Society for generously funding this research with an Exploration Grant. We also thank the following experts for participating in the evaluation: Jürg Gilgen, Thomas Wehrli and Hans-Uli Feldmann (swisstopo), Martin Gamache and Alex Tait (National Geographic Society), Tom Patterson (U.S. National Park Service), Ernst Spiess and Stefan Räber (ETH Zurich), Patrick Kennelly (Central Oregon Community College), Matthias Beilstein (beilstein.biz), Claudia A. Trochsler (CAT Design), Michael Wood (University of Aberdeen), Roger Smith (Geographx Ltd), Arne Rohweder (rohweder-map-design.com), Markus Hauser (Orell Füssli Kartographie AG), Sarah Bell (Esri Inc.), Olga Kovaleva (Moscow State University of Geodesy and Cartography), and Brooke Marston (with thanks for editing this text). And finally we thank the reviewers for their valuable comments.

# Cartographic Relief Shading with Neural Networks

Bernhard Jenny, Magnus Heitzler, Dilpreet Singh, Marianna Farmakis-Serebryakova, Jeffery Chieh Liu and Lorenz Hurni

Supplementary Material: Shaded relief images for expert evaluation

———————————◆———————————

Figures 1 to 10 below show the ten pairs of manual and neural network shaded relief images that were used for the expert evaluation. For this online survey, each pair of images was displayed with a total width of 1,000 pixels. All pairs were arranged horizontally, except #1 *Tutuila*, which was arranged vertically. The numbering of figures is identical to the numbering of pairs in the paper. Note that the order of manual and neural network shadings varies among pairs as in the survey.

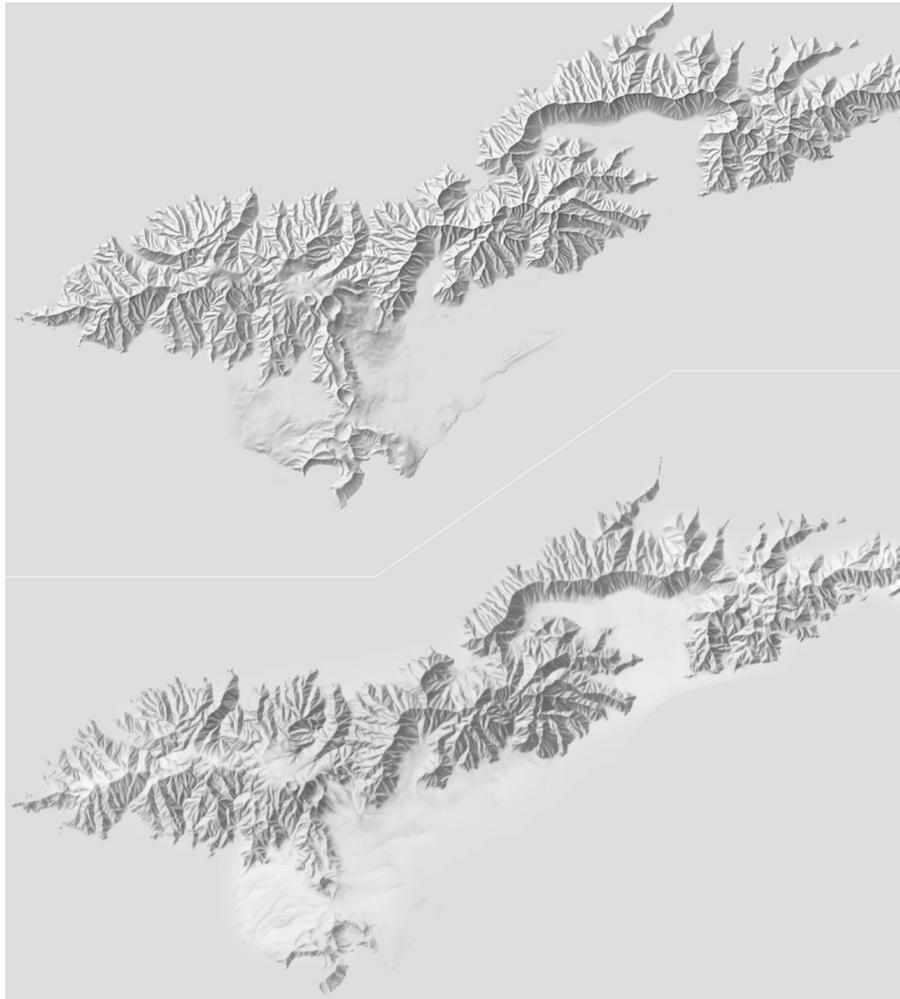

Fig. 1. #1 Tutuila, American Samoa, neural network (top) and manual shading (bottom). Manual shading by Michael Wood, University of Aberdeen, Scotland, UK, available at shadedreliefarchive.com, and published in J. P. Theroux and E. A. Wingert, *A Coastal zone management atlas of American Samoa*. University of Hawaii Cartographic Laboratory, Department of Geography, 1981.

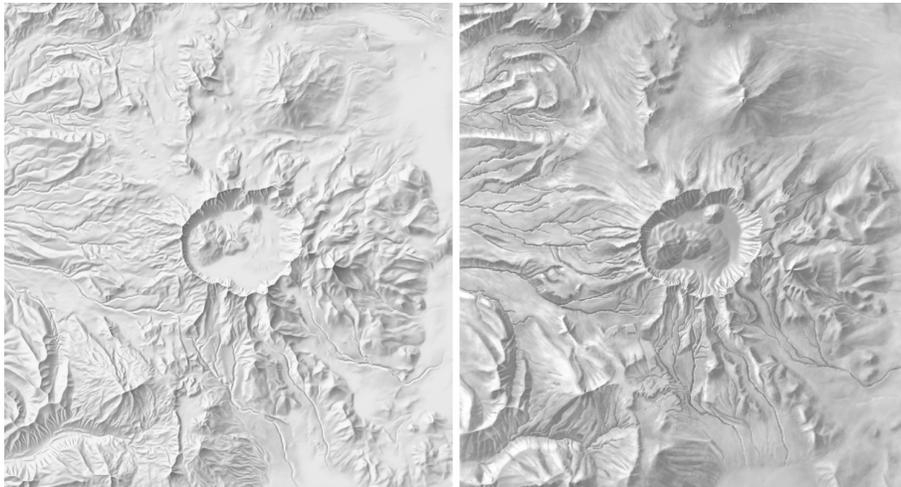

Fig. 2. #2 Crater Lake National Park, Oregon, USA, neural network (left) and manual shading (right). Manual shading by Bill von Allmen, U.S. National Park Service, available at shadedreliefarchive.com.

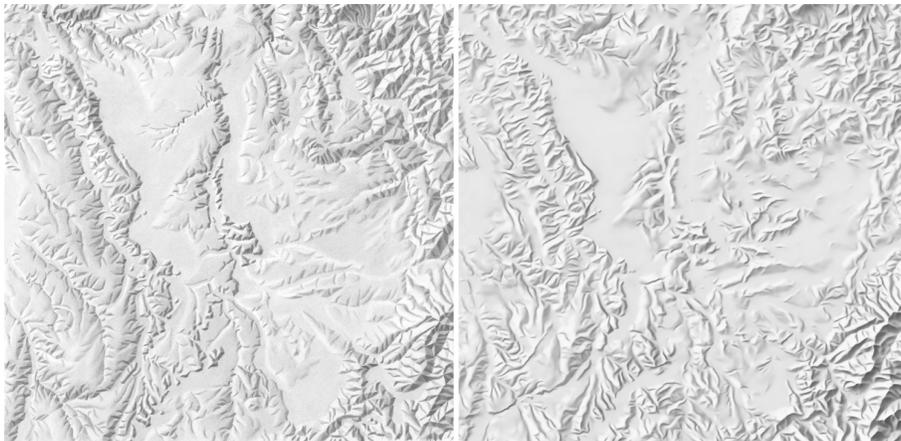

Fig. 3. #3 Lorraine, manual (left) and neural network shading (right). Manual shading from Swiss World Atlas, Schweizer Weltatlas – Atlas Mondial Suisse – Atlante Mondiale Svizzero. EDK Schweizerische Konferenz der kantonalen Erziehungsdirektoren, Lehrmittelverlag Zürich, 2017. © Swiss Conference of Cantonal Ministers of Education (EDK).

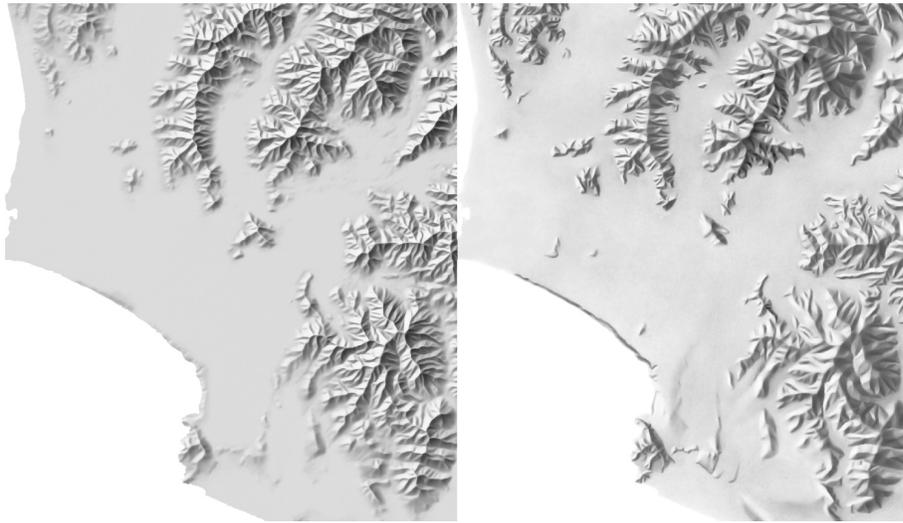

Fig. 4. #4 Lima, Peru, neural network (left) and manual shading (right). Manual shading from Swiss World Atlas, Schweizer Weltatlas – Atlas Mondial Suisse – Atlante Mondiale Svizzero. EDK Schweizerische Konferenz der kantonalen Erziehungsdirektoren, Lehrmittelverlag Zürich, 2017. © Swiss Conference of Cantonal Ministers of Education (EDK).

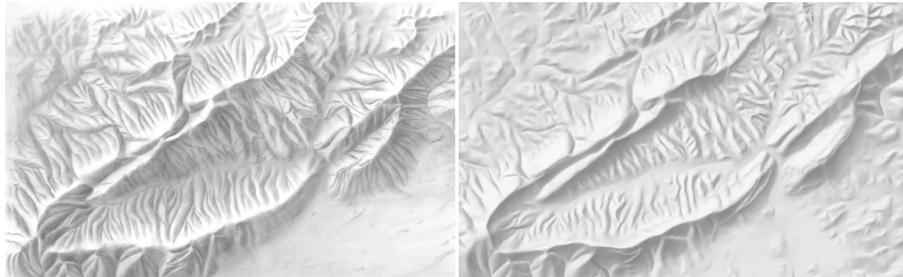

Fig. 5. #5 Hot Springs National Park, Arkansas, USA, manual (left) and neural network shading (right). Manual shading by U.S. National Park Service, available at shadedreliefarchive.com.

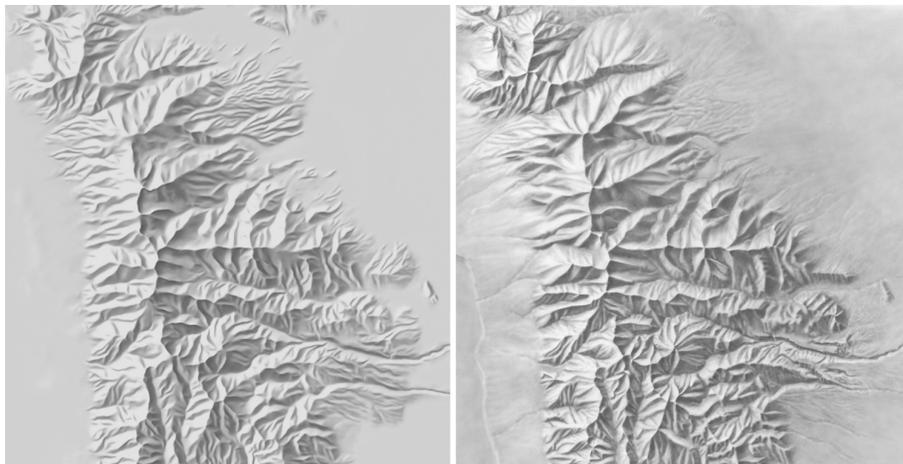

Fig. 6. #6 Great Basin National Park, Nevada, USA, neural network (left) and manual shading (right). Manual shading by Bill von Allmen, U.S. National Park Service, available at shadedreliefarchive.com.

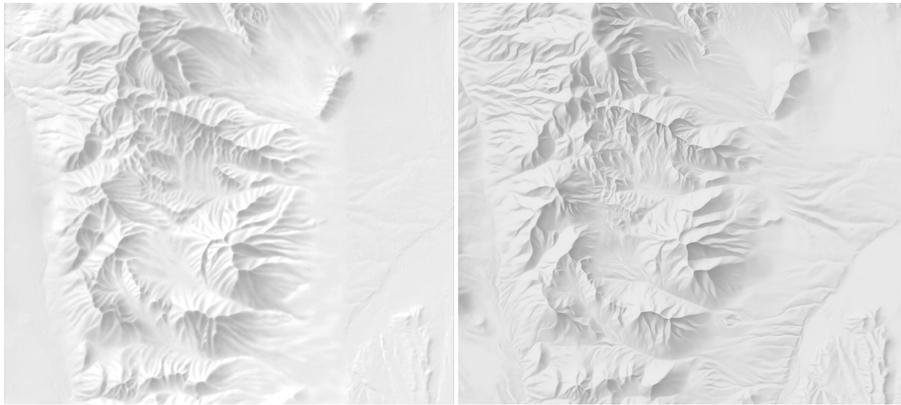

Fig. 7. #7 City of Rocks National Reserve, Idaho, USA, manual (left) and neural network shading (right). Manual shading by Tom Patterson, U.S. National Park Service, available at shadedreliefarchive.com.

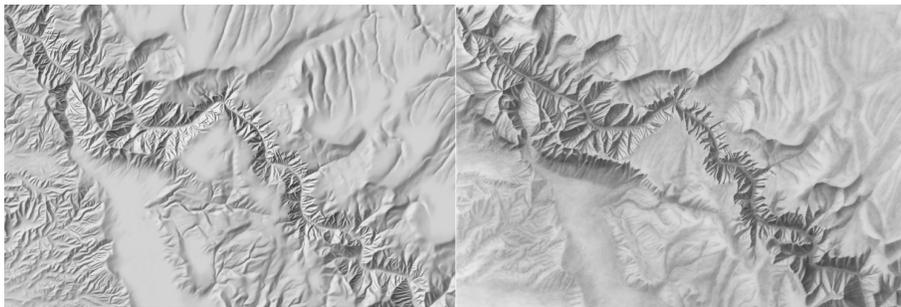

Fig. 8. #8 Black Canyon National Park, Colorado, USA, neural network shading (left) and manual shading (right). Manual shading by Bill von Allmen, U.S. National Park Service, available at shadedreliefarchive.com.

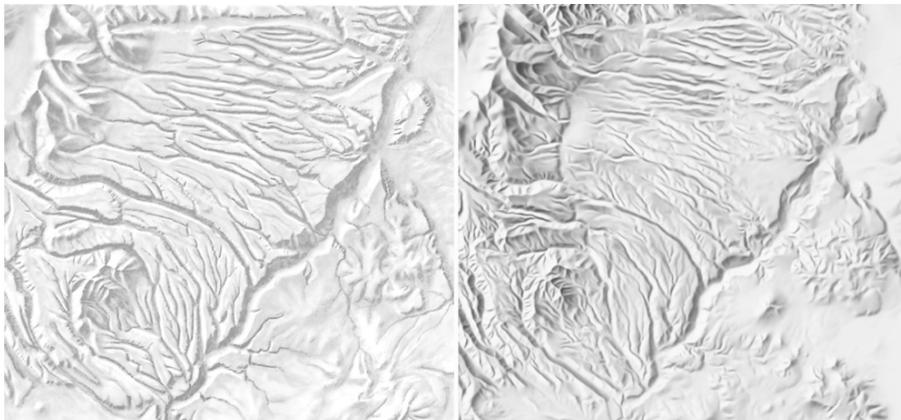

Fig. 9. #9 Bandelier National Monument, New Mexico, USA, manual (left) and neural network shading (right). Manual shading by Bill von Allmen, U.S. National Park Service, available at shadedreliefarchive.com.

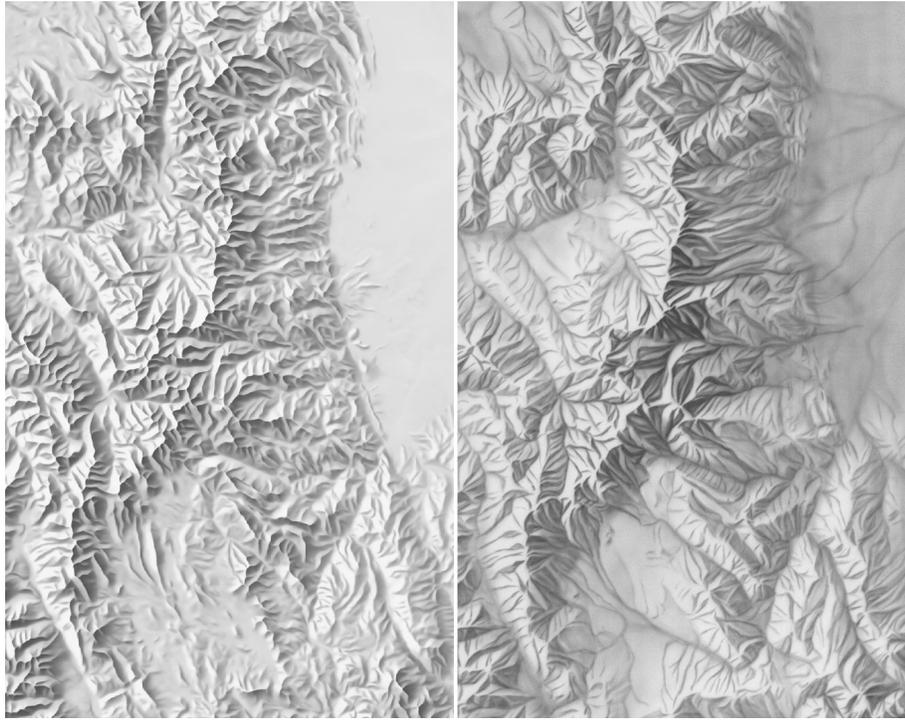

Fig. 10. #10 Rocky Mountain National Park, Colorado, USA, neural network (left) and manual shading (right). Manual shading by Bill von Allmen, U.S. National Park Service, available at shadedreliefarchive.com.